# Gold-Patched Graphene Nanoribbons for High-Responsivity and Ultrafast Photodetection from Visible to Infrared Regimes

Semih Cakmakyapan*, Ping Keng Lu, Aryan Navabi, and Mona Jarrahi*

Electrical Engineering Department, University of California Los Angeles

Email addresses: (1) semih@ucla.edu, (2) mjarrahi@ucla.edu

## Abstract

Graphene is a very attractive material for broadband photodetection in hyperspectral imaging and sensing systems. However, its potential use has been hindered by tradeoffs between the responsivity, bandwidth, and operation speed of existing graphene photodetectors. Here, we present engineered photoconductive nanostructures based on gold-patched graphene nanoribbons, which enable simultaneous broadband and ultrafast photodetection with high responsivity. These nanostructures merge the advantages of broadband optical absorption, ultrafast photocarrier transport, and carrier multiplication in graphene nanoribbons with the ultrafast transport of photocarriers to the gold patches before recombination. Through this approach, high-responsivity operation is achieved without the use of bandwidth- and speed-limiting quantum dots, defect states, or tunneling barriers. We demonstrate high-responsivity photodetection from the visible to the infrared regime (0.6 A/W at 0.8 µm and 11.5 A/W at 20 µm) with operation speeds exceeding 50 GHz. Our results demonstrate an improvement of the response times by more than seven orders of magnitude and an increase in bandwidths of one order of magnitude compared to those of higher-responsivity graphene photodetectors based on quantum dots and tunneling barriers.

## Main

Graphene has rapidly become an attractive candidate material for broadband and ultrafast photodetection because of its distinct optical and electronic characteristics [1, 2]. These characteristics stem from the unique band structure of graphene, which allows carrier generation by optical absorption over an extremely broad spectral range from the ultraviolet to the microwave regime. Moreover, the high electron/hole mobility and weak scattering in graphene enable ultrafast temporal responses in graphene photodetectors. Additionally, the two-dimensional nature of graphene enables the generation of multiple electron/hole pairs for a single absorbed photon [3-5]. Furthermore, the compatibility of graphene photodetectors with silicon-based fabrication platforms enables their integration with low-cost and high-performance complementary metal oxide semiconductor (CMOS) read-out and post-processing circuits.

Most graphene photodetectors utilize graphene–metal junctions or graphene p–n junctions to spatially separate and extract the photogenerated carriers. A low optical absorption inside the effective junction regions (~100−200 nm) and short photocarrier lifetime of graphene (~1 ps) have been the two major challenges for developing high-responsivity graphene photodetectors [6]. Various techniques have been explored to address these challenges and to enhance the responsivity of graphene photodetectors. Despite the significant benefits of these techniques for offering high photodetection responsivities, the scope and potential use of existing graphene photodetectors remain limited by the tradeoffs between their high responsivity, ultrafast temporal responses, and broadband operation.

For example, responsivity-enhanced photodetection from visible to mid-infrared wavelengths has been achieved by increasing the photocarrier lifetime through band-structure engineering and defect engineering in graphene. For this purpose, carrier trapping mechanisms and patterned graphene nanostructures have been used to introduce bandgap and midgap defect states in graphene [7-10]. However, the response times of these graphene photodetectors have been limited by long carrier trapping times in the introduced defect and edge states. Hybrid graphene–quantum dot photodetectors have been a powerful alternative for enhancing photodetection responsivity by increasing light absorption and introducing large carrier multiplication factors [11-14]. However, the bandwidth and response time of this type of graphene photodetectors have been restricted by the narrow spectral bandwidth of the quantum dots and long carrier trapping times in the quantum dots. Photodetectors based on two graphene layers separated by a thin tunnel barrier have also offered enhanced broadband responsivity by separating the photogenerated electrons and holes through quantum tunneling and minimizing their recombination [15]. However, the response times of this type of graphene photodetectors have been limited by the long carrier trapping times in the utilized tunneling barriers. Waveguide-integrated graphene photodetectors have been another promising alternative for offering enhanced ultrafast responsivity by increasing the interaction length of light within graphene [16-19]. These graphene photodetectors have the additional advantage of process-compatibility with standard photonic integrated circuits. However, their spectral bandwidth has been restricted by the bandwidth limitations of the utilized waveguides. Moreover, microcavities, plasmonic structures, and optical antennas have been integrated with graphene to achieve high responsivities by increasing the interaction length of light within graphene [20-32]. However, the bandwidth of these types of graphene photodetectors have been limited by the resonant nature of the utilized structures.

In this work, we use engineered photoconductive nanostructures based on gold-patched graphene nanoribbons, which offer unique electrical and optical characteristics that enable simultaneous broadband and ultrafast photodetection with high responsivity. The key novelty that enables the superior performance of these photoconductive nanostructures is that they constrain most of the photocarrier generation and conduction to the graphene nanoribbons and gold patches, respectively. Therefore, they benefit from the broadband optical absorption and photocarrier multiplication capabilities of graphene while avoiding the negative effects of the short photocarrier lifetime of graphene.

The operation principle of the photodetector based on the utilized photoconductive nanostructures is illustrated in Fig. 1a. Arrays of monolayer graphene nanoribbons are patterned on a high-resistivity Si substrate (> 10 kΩ.cm) coated with a 130 nm thick thermal oxide. The carrier concentration and sheet resistance of the graphene nanoribbons used in this study are measured to be $1.8 \times 10^{13}$ cm$^{-2}$ and 804 Ω/□ when no gate bias is applied (Supplementary Fig. S1). The photoconductive nanostructures are formed by connecting arrays of nanoscale gold patches to either side of the graphene nanoribbons. The geometry of the gold patches is engineered to concentrate a major portion of the incident optical beam onto the graphene nanoribbons over a broad optical spectrum ranging from the visible to infrared regimes. The graphene nanoribbons are designed to be narrower than the effective metal-graphene junction regions, where the photogenerated electron and holes separate. This design enables a fast photocarrier transit time to the gold patches under an applied bias voltage, and this transit time is much faster than the graphene photocarrier lifetime. Therefore, unlike the previously demonstrated graphene photodetectors, which achieved high photoconductive gains via an increase of the photocarrier lifetime, the gold-patched graphene nanoribbons offer high photoconductive gains by reducing the photocarrier transport time to the gold patches. The photocarriers transported to the gold patches are all combined (illustrated by red arrows in Fig. 1a) to form the output photocurrent of the photodetector.

Numerical finite difference time domain simulations (Lumerical) are carried out to demonstrate the unique capability of the designed gold patches for efficiently concentrating the incident optical beam onto the graphene nanoribbons over a broad optical wavelength range. Figure 1b shows the color plot of the transmitted optical field through the gold patches at 0.8 μm, 5 μm, and 20 μm for an incident optical beam polarized normal to the graphene nanoribbons, indicating highly efficient and broadband optical coupling to the graphene nanoribbons. The numerical analysis predicts that 40%-60% of the incident optical beam would be focused onto the graphene nanoribbons over a broad spectrum ranging from the visible to infrared regimes, as illustrated in Fig. 1c (red curve). Despite the broadband optical coupling to the graphene nanoribbons, the optical absorption in graphene is estimated to be highly wavelength dependent, as illustrated in Fig. 1c (blue curve). This strong wavelength dependence stems from the optical absorption in graphene being dominated by interband transitions in the visible and near-infrared spectral ranges and by intraband transitions in the infrared spectral range [33], leading to much lower optical absorption in the visible and near-infrared regimes.

One of the unique features of the gold-patched graphene nanoribbons is that they exploit enhanced carrier multiexcitation generation at higher photon energy levels to compensate for the lower optical absorption at lower wavelengths [3-5]. Such carrier multiplication factors have not been previously exploited in monolayer graphene photodetectors without the use of quantum dots because of the short photocarrier lifetimes in graphene [14]. However, they can be used to boost the photoconductive gain of the gold-patched graphene nanoribbons at lower wavelengths because of the fast photocarrier transport time to the gold patches. Because the use of any defect states and/or quantum dots is avoided, the utilized gold-patched graphene nanoribbons enable high responsivity photodetection without sacrificing the broadband and ultrafast operation.

Figure 1d shows an optical microscope image of a fabricated photodetector based on the gold-patched graphene nanoribbons; the photodetector has an active area of 30 × 30 $\mu m^2$. This figure also shows a scanning electron microscopy (SEM) image of the utilized gold patches (for further details, see Methods). A supercontinuum laser and a Globar light source combined with bandpass filters are used to measure the photodetector responsivity in the visible/near-infrared and infrared regimes, respectively (for further details, see Methods). The responsivity spectrum of the fabricated photodetector at an optical power of 2.5 μW, a gate voltage of 25 V, and a bias voltage of 20 mV is shown in Fig. 1e (red data). The photodetector offers an ultrabroad operation bandwidth from the visible to the infrared regime with high-responsivity levels ranging from 0.6 A/W (at an 800 nm wavelength) to 8 A/W (at a 20 μm wavelength). This graphene photodetector exhibits the widest photodetection bandwidth with high responsivity reported to date, which was enabled by the use of gold-patched graphene nanoribbons. In the following sections, we explain that even higher responsivity levels can be achieved by the presented photodetector when the bias and gate voltages are optimized. In addition, the asymmetric geometry of the utilized gold-patched graphene nanoribbons leads to a highly polarization-sensitive responsivity (Supplementary Fig. S2). The strong polarization sensitivity of the presented photodetector could find many applications in polarimetric imaging and sensing systems.

The photoconductive gain of the fabricated photodetector is calculated from the measured responsivity and the estimated optical absorption in graphene, as illustrated in Fig. 1e (blue data). As expected, higher photoconductive gains are achieved at lower wavelengths, at which the photogenerated electrons are excited to higher energy levels in the conduction band. Excitation to higher energy levels gives rise to the excitation of secondary electron-hole pairs by transferring more energy during relaxation, as illustrated in the inset of Fig. 1e.

Figure 2 shows the effect of the gate voltage on the photodetector responsivity in the visible and near-infrared regimes, where the optical absorption is dominated by interband transitions in the graphene. The effect of the gate voltage is best described by the device band diagrams at various gate voltages (Fig. 2a inset), which illustrate that the gate voltage tunes the carrier concentration and Fermi energy level of the graphene nanoribbons between the gold patches while maintaining the same metal-induced doping levels at the gold-patch junctions. This tuning modifies the band bending slope at different gate voltages (A detailed analysis of the device band diagram is included in the Supplementary Fig. S3). When an optical beam is incident on the device, the photogenerated electrons and holes move according to the induced electric field determined by the band bending slope. Therefore, the photogenerated holes move to the center of the graphene nanoribbons, where they eventually recombine, and the photogenerated electrons move to the anode and cathode junctions. The induced photocurrent is proportional to the difference of the photogenerated electrons that drift to the anode and cathode junctions (Supplementary Fig. S4).

Figure 2a shows the responsivity of the fabricated photodetector at a wavelength of 800 nm and a bias voltage of 20 mV when the gate voltage is varied between −20 V and 65 V. As the gate voltage decreases from 65 V, the p-type carrier concentration in the graphene nanoribbons increases and a steeper energy band bending is introduced on the anode side than the cathode side [34, 35]. The variations in the band bending slopes lead to an increase in the induced photocurrent when decreasing the gate voltage. This trend changes at the gate voltages less than 22 V and a small decrease in the induced photocurrent is observed when decreasing the gate voltage further. The observed decrease in photocurrent at the gate voltages less than 22 V can be explained by the reduction of carrier mobility at high carrier densities because of the increase in carrier scattering [36-38]. The highest responsivity level is achieved at a gate voltage of 22 V. Because carrier multiplication is an important factor that contributes to the high photoconductive gain of the photodetector in the visible and near-infrared regimes, the photodetector responsivity decreases at higher optical powers (Supplementary Fig. S5). This decrease is due to the increase in the carrier recombination rate at high photogenerated carrier densities, which reduces the carrier scattering time and the carrier multiplication efficiency [3-5]. Larger area gold-patched graphene nanoribbon arrays can be used to maintain high photo-detection efficiencies at high optical powers. As expected, the measured photodetector responsivity has a linear dependence on the applied bias voltage (Supplementary Fig. S6). This dependence suggests that higher responsivity levels can be achieved by increasing the bias voltage at the expense of an increased dark current.

Figure 2b indicates how the gate voltage affects the photodetector responsivity by varying the optical absorption of graphene. The measured responsivity values in the visible and near-infrared regimes at gate voltages of -20 V, 22 V, and 65 V are shown. The responsivity values at each gate voltage are divided by the photodetector responsivity at 800 nm to eliminate the influence of variations in the band diagram at different gate voltages. Since the photodetector responsivity is proportional to the optical absorption coefficient divided by the photon energy, as far as optical absorption is governed by interband transitions in graphene, the photodetector responsivity should be linearly proportional to the photon wavelength (dashed line: Responsivity/Responsivity at 0.8 μm = λ/0.8 μm). However, because interband optical transitions are allowed when $\hbar\omega \geq 2E_f$, the photodetector responsivity is substantially reduced by Pauli blocking at lower photon energies when the gate voltage is lowered from 65 V to 22 V and -20 V.

Figure 3 shows the impact of the gate voltage on the photodetector performance in the infrared regime, in which the optical absorption is dominated by intraband transitions in the graphene. The impact of the gate voltage is best described by the graphene band diagrams at various gate voltages and wavelengths (Fig. 3a inset). They illustrate that the gate voltage tunes the Fermi energy level, which changes the number of available states because of the cone-shaped band diagram of graphene. Because a larger number of states is available at higher Fermi energies, the photodetector responsivity values are increased by a decrease in the gate voltage at all wavelengths. At a given Fermi energy level, a larger number of states is available to be filled by lower-energy photons, which results in an increase in photodetector responsivity values at longer wavelengths. Responsivity values as high as 11.5 A/W are achieved at a 20 μm wavelength and -20 V gate voltage, which corresponds to the lowest photon energy and highest Fermi energy level in our measurements, respectively. Notably, the operation bandwidth of the presented photodetector is not limited to 20 μm, and higher responsivity values are expected at longer wavelengths. The measurement bandwidth of our experimental setup was limited by the detection bandwidth of the calibrated infrared detector used for the measurements (for further details, see Methods).

One of the drawbacks of the presented photodetector based on gold-patched graphene nanoribbons is its relatively large dark current due to the photoconductive nature of the photodetector. Therefore, the noise equivalent power (NEP) of the fabricated photodetector is calculated to assess the noise performance. For this calculation, we assume an optical chopping frequency above 1 kHz and lock-in detection of the output signal to significantly reduce the $1/f$ noise current with respect to the Johnson Nyquist and shot-noise currents, the same way commercially available room-temperature infrared detectors are operated [39, 40]. Under this operation condition, the photodetector noise current, which is dominated by the Johnson Nyquist and shot-noise sources, is extracted from the measured photocurrent and resistance data (supplementary Fig. S7). The calculated NEP levels from the extracted noise current and measured responsivity values are in the range from 1 to 10 pW/Hz$^{1/2}$ (Fig. 3b), exhibiting superior noise performance compared to commercially available room-temperature infrared detectors under similar operation conditions [39, 40].

A unique attribute of the presented graphene photodetector is that its superior bandwidth/responsivity performance is accompanied by an ultrafast photodetection speed. This ultrafast speed became possible through a special design of the utilized gold-patched graphene nanoribbons, which offers broadband optical absorption in the graphene and subpicosecond photocarrier transport times to the gold patches while maintaining low capacitive/resistive parasitics. A high-frequency electrical model of the graphene photodetector is shown in Fig. 4. The graphene resistance, $R_g$, and $SiO_2$ capacitance, $C_{ox}$, are measured to be 70 Ω and 5.2 pF, respectively. Additionally, the gold patch capacitance, $C_g$, and substrate resistance, $R_{sub}$, are estimated as 12.9 fF and 5 MΩ, respectively. As it can be observed from the device electrical model, by use of a high-resistivity silicon substrate, a large resistance, $R_{sub}$, is placed in series with the $SiO_2$ capacitance, $C_{ox}$, eliminating the negative impact of this capacitance on the ultrafast photodetection speed. Therefore, the photodetector frequency response predicted by this electrical model is dominated by the parasitic resistance of the graphene nanoribbons and capacitive parasitic of the gold patches, leading to a predicted photoresponse cutoff frequency of 425 GHz.

The operation speed of the fabricated photodetector is characterized using two fiber-coupled, wavelength-tunable, distributed-feedback (DFB) lasers with 783 nm and 785 nm center wavelengths (TOPTICA #LD-0783-0080-DFB and #LD-0785-0080-DFB), as illustrated in Fig. 4. Both lasers have a 2 MHz spectral linewidth and 2.4 nm wavelength tunability range. When the two laser beams are combined in a single-mode fiber, they provide a tunable optical beating frequency ranging from 20 MHz to 2 THz. The combined laser beams are used to illuminate the fabricated graphene photodetector to induce a photocurrent at the optical beating frequency. A small portion of the optical beam is monitored through an optical spectrum analyzer to ensure that the two beating optical beams are maintained at the same power level. The photodetector output is probed by a GSG microwave probe (Picoprobe 50A-GSG-100-P-N) and monitored with a spectrum analyzer (HP8566B). A broadband bias-T (HP11612B) is used to apply the bias voltage and to out-couple the high-frequency a.c. photocurrent. Figure 4 shows the photodetector output at various beating frequencies. As expected from the theoretical predictions, the experimental results exhibit no roll-off in the photodetector response up to 50 GHz, which is the frequency limitation of the utilized GSG probes and spectrum analyzer. The variations in the photodetector output are due to the fluctuations in the output power of the DFB lasers.

Figure 5 compares the specifications of the presented graphene photodetector based on gold-patched graphene nanoribbons with the specifications of some of the highest-performance room-temperature graphene photodetectors reported in the literature [11, 14, 15, 16, 17, 19, 29, 41, 42]. The presented graphene photodetector offers responsivity levels more than two orders of magnitude higher than those of previously reported high-speed graphene photodetectors [16, 17, 19]. It also offers response times more than seven orders of magnitude faster and bandwidths one order of magnitude broader compared to those of the higher-responsivity graphene photodetectors based on quantum dots [11, 14] and tunneling barriers [15]. We expect this unique combination of broadband and ultrafast photodetection with high responsivity enabled by the gold-patched graphene nanoribbons will have a significant impact on future hyperspectral imaging and sensing systems. To further enhance the device performance, the symmetric gold patches could be replaced with asymmetric metal patches, which would break the symmetry of the electrical potential inside the graphene nanoribbons and allow bias-free, low-dark-current device operation [43].

## Methods

**Device Fabrication:** Commercially available chemical vapor deposition (CVD)-grown graphene is first transferred to a high-resistivity silicon wafer covered with a 130 nm-thick thermally grown $SiO_2$ layer. The gold patches are then patterned by electron beam lithography and formed by 5/45 nm Ti/Au deposition and liftoff. The graphene nanoribbons are then patterned by another electron beam lithography step and formed by oxygen plasma etching. Next, the bias lines and output pads are formed by optical lithography and formed by 20/200 nm Ti/Au deposition and liftoff. Finally, the gate pads are patterned by optical lithography and formed by $SiO_2$ plasma etching, followed by 20/200 nm Ti/Au deposition and liftoff.

**Responsivity Measurements:** A supercontinuum laser (NKT Photonics - SuperK EXTREME) is used to measure the photodetector responsivity in the visible/near-infrared regime (0.8-1.8 μm wavelength range). The fiber-coupled output of the supercontinuum laser is placed very close to the device to ensure that all output power is incident on the 30 × 30 $\mu m^2$ device active area. The responsivity values are calculated from the measured photocurrent using a source measure unit instrument (Keithley - 2450

SourceMeter) and the measured optical power using a calibrated near-infrared photodetector (Thorlabs - S132C). A Globar light source (Thorlabs - SLS203L) combined with different infrared bandpass filters is used to measure the photodetector responsivity in the infrared range (3-20 μm wavelength range). A list of the filters used and their spectral characteristics are included in supplementary Table S1. A calibrated calorimeter (Scientech - AC2500S Calorimeter and Vector S310) is used to measure the infrared radiation at each wavelength. The calorimeter is positioned 1 cm from the Globar output, where the infrared intensity is uniform across the calorimeter input aperture. The uniformity of the infrared beam is confirmed by replacing the calorimeter with a graphene photodetector and monitoring its output photocurrent while moving it in the plane normal to the incident beam. The responsivity values are calculated from the measured photocurrent and the measured infrared power using the calorimeter and scaled by the ratio between the active area of the graphene photodetector and the calorimeter.

## Data Availability

The data that support the findings of this study are available from the corresponding authors upon request

## Acknowledgments

The authors gratefully acknowledge the financial support from the Department of Energy (grant # DE-SC0016925) and thank Dr. Ning Wang, Dr. Sergei Tochitsky, and Bor-Chau Juang for their helpful suggestions during device characterization.

## Author Information

### Affiliations

University of California Los Angeles Electrical Engineering Department, 420 Westwood Plaza, Los Angeles, CA, 90025, USA

## Contributions

S. C. designed, fabricated, and characterized the device prototypes. P. K. L. and A. N. assisted with the device fabrication. S. C. and M. J. wrote the manuscript with inputs from the co-authors. M. J. supervised the work.

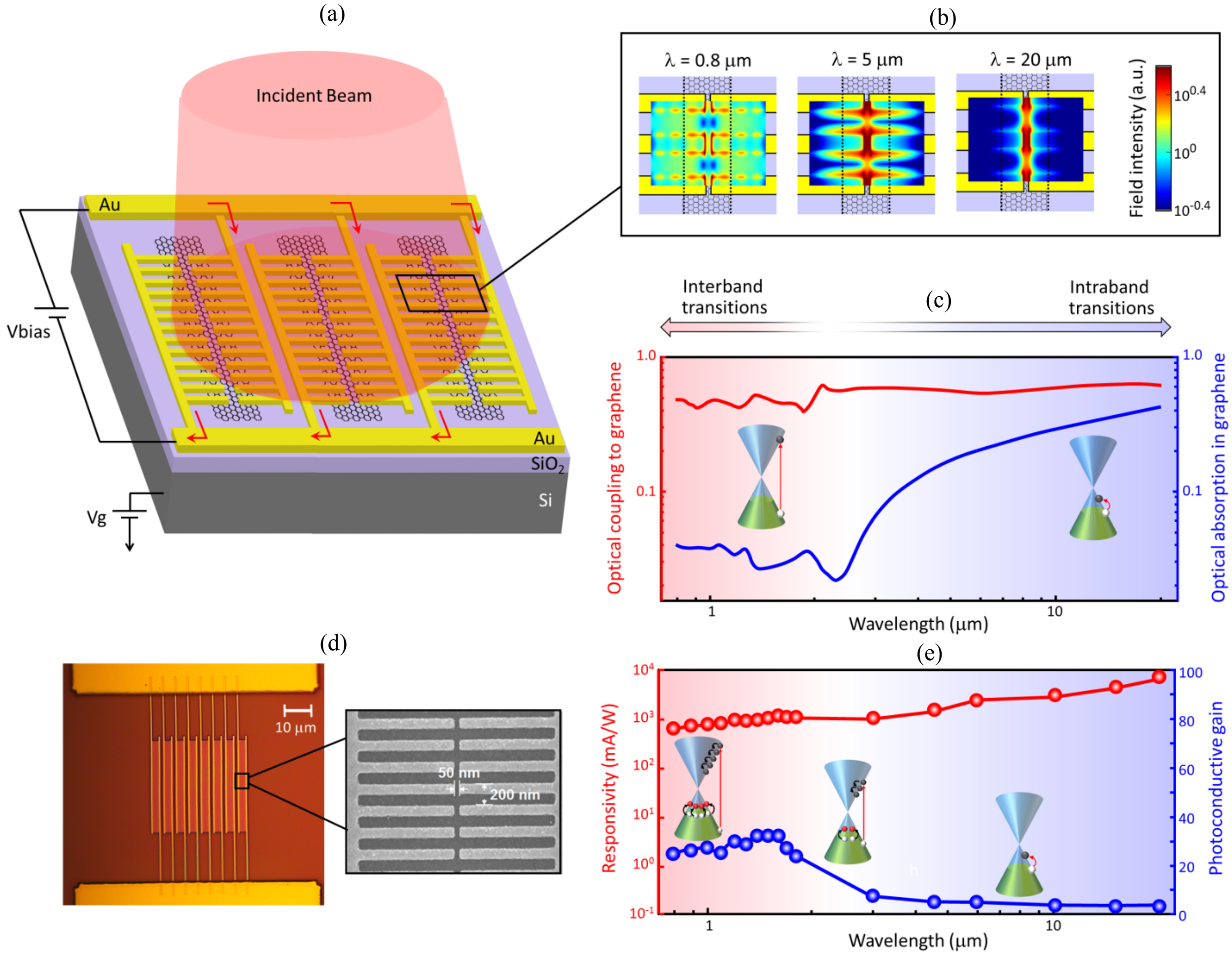

Fig. 1. High-responsivity and broadband photodetection through gold-patched graphene nanoribbons. (a) Schematic of a photodetector based on gold-patched graphene nanoribbons. It is fabricated on a high-resistivity Si substrate coated with a 130 nm thick $SiO_2$ layer. The gate voltage applied to the Si substrate, $V_g$, controls the Fermi energy level of the graphene nanoribbons. The gold patches have a width of 100 nm, a periodicity of 200 nm, a height of 50 nm, a length of 1 μm, and a tip-to-tip gap size of 50 nm. (b) Color plot of the transmitted optical field, polarized normal to the graphene nanoribbons, through the gold patches at 0.8 μm, 5 μm, and 20 μm, indicating highly efficient and broadband optical coupling to the graphene nanoribbons. (c) Numerical estimates of the optical coupling (red curve) and optical absorption (blue curve) in the graphene nanoribbons as a function of the wavelength. (d) Optical microscope and SEM images of a fabricated photodetector based on the gold-patched graphene nanoribbons. (e) The measured responsivity (red data) and photoconductive gain (blue data) of the fabricated photodetector at an optical power of 2.5 μW, gate voltage of 22 V, and bias voltage of 20 mV.

(a)

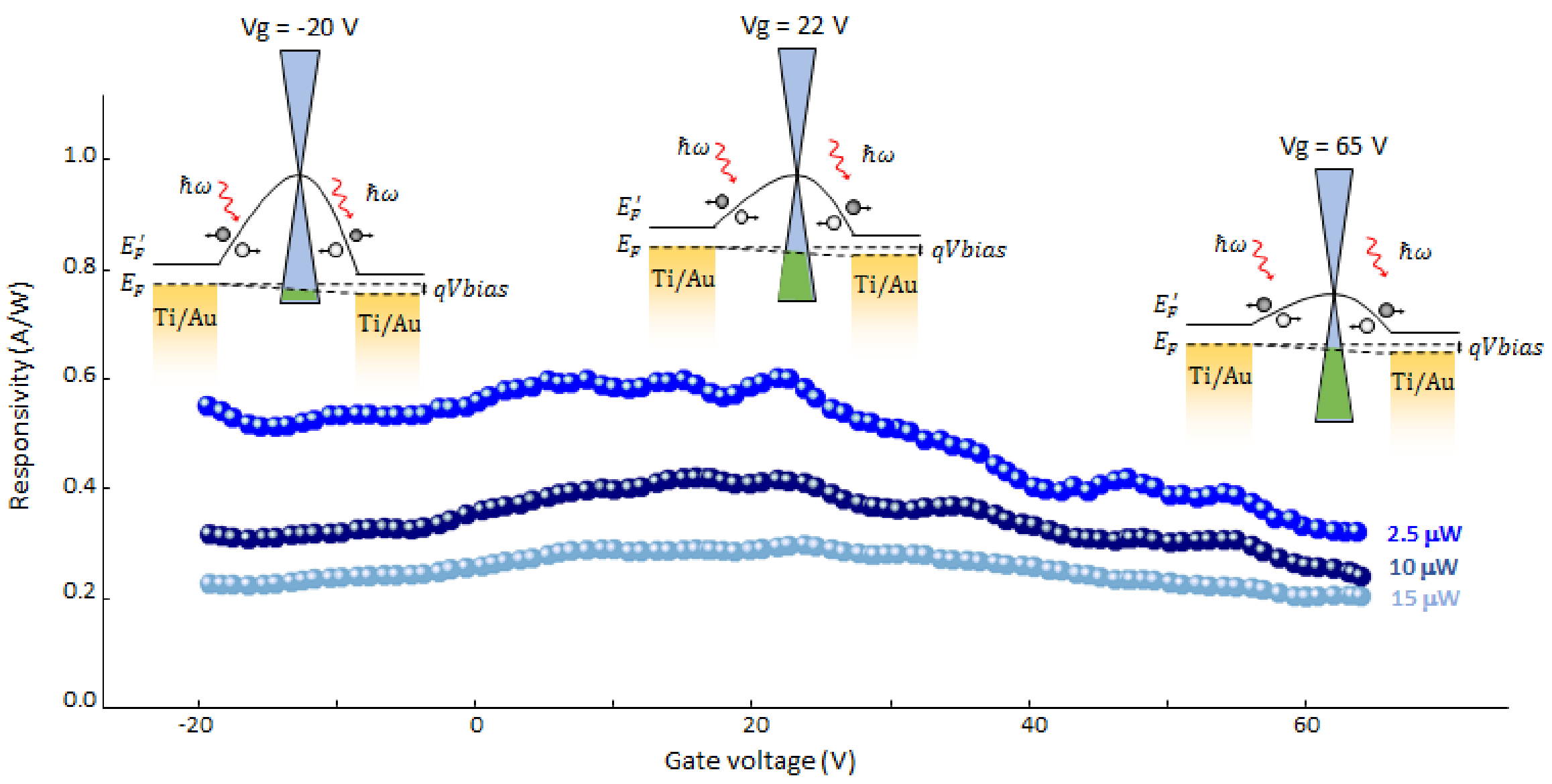

(b)

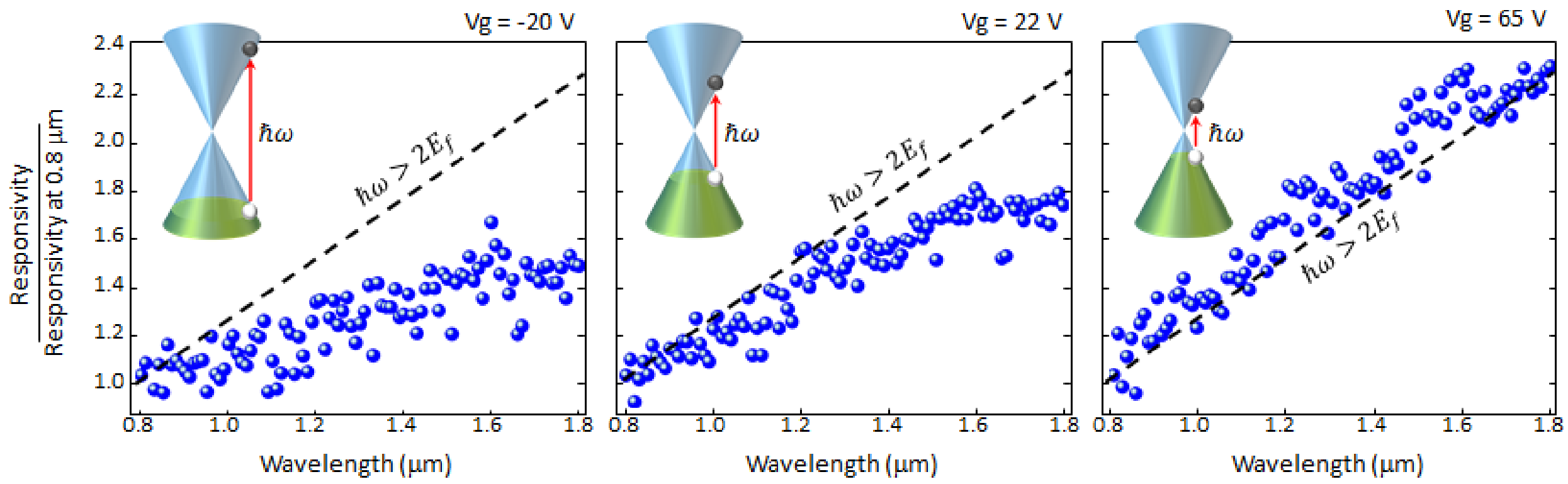

Fig. 2. Impact of the gate voltage on the photodetector performance in the visible and near-infrared regimes. (a) Responsivity of the fabricated photodetector at an 800 nm wavelength under a bias voltage of 20 mV. Inset figures show the band diagram of the graphene photodetector at gate voltages of -20 V, 22 V, and 65 V. The Fermi energy level ($E_F$) and the graphene Dirac point energy level ($E_F$`) are illustrated by the dashed and solid black lines, respectively. The carrier concentration of the graphene nanoribbons used in this study is measured to be $1.8 \times 10^{13}$ cm$^{-2}$ when the gate is unbiased, indicating highly p-doped graphene nanoribbons at a gate voltage of -20 V. (b) Responsivity of the fabricated photodetector at optical wavelengths ranging from 800 nm to 1.8 μm and at gate voltages of -20 V, 22 V, and 65 V. The responsivity values at each gate voltage are divided by the photodetector responsivity at 800 nm to eliminate the influence of variations in the band diagram at different gate voltages. The dashed lines show the predicted responsivity spectra if interband transitions are allowed over the entire wavelength range.

(a)

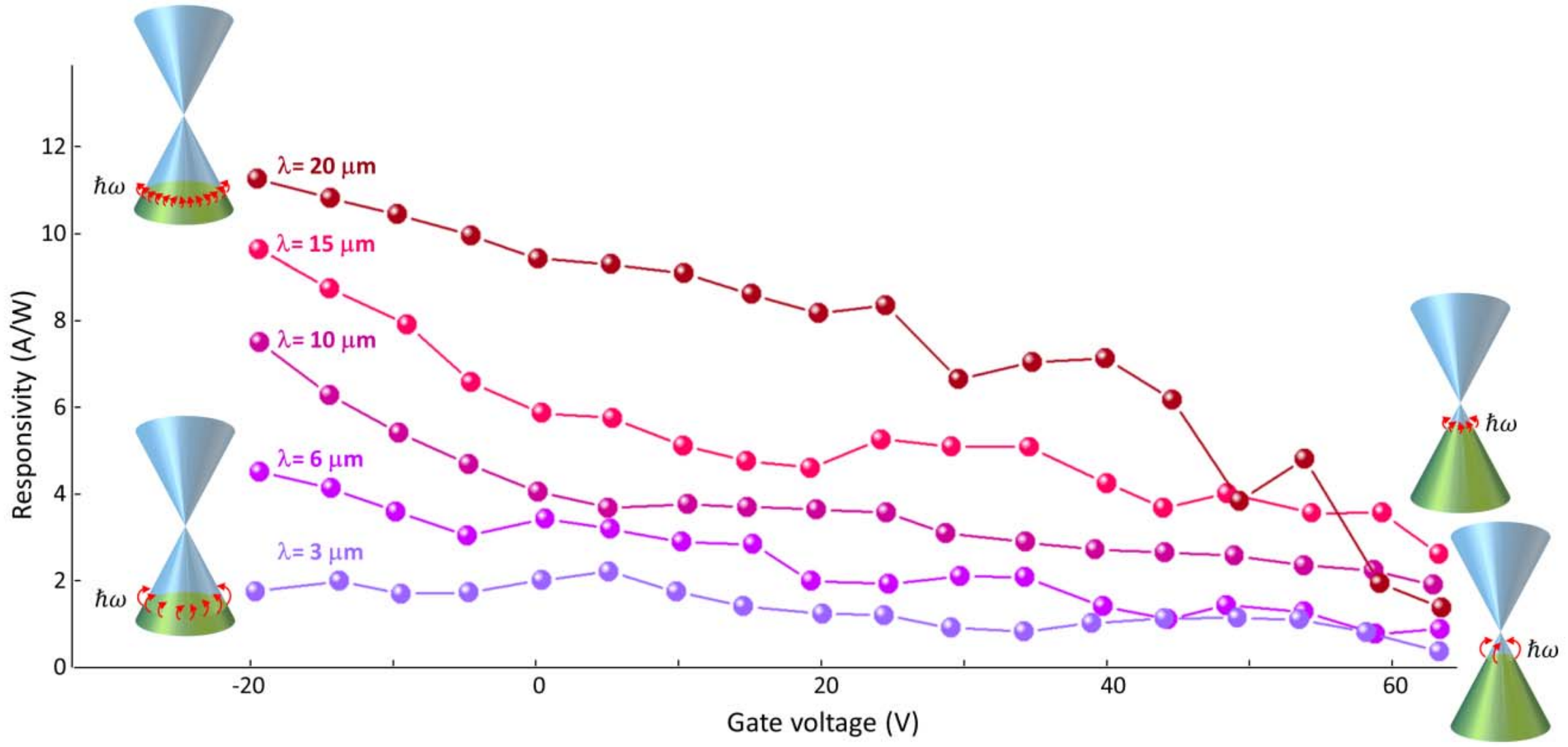

(b)

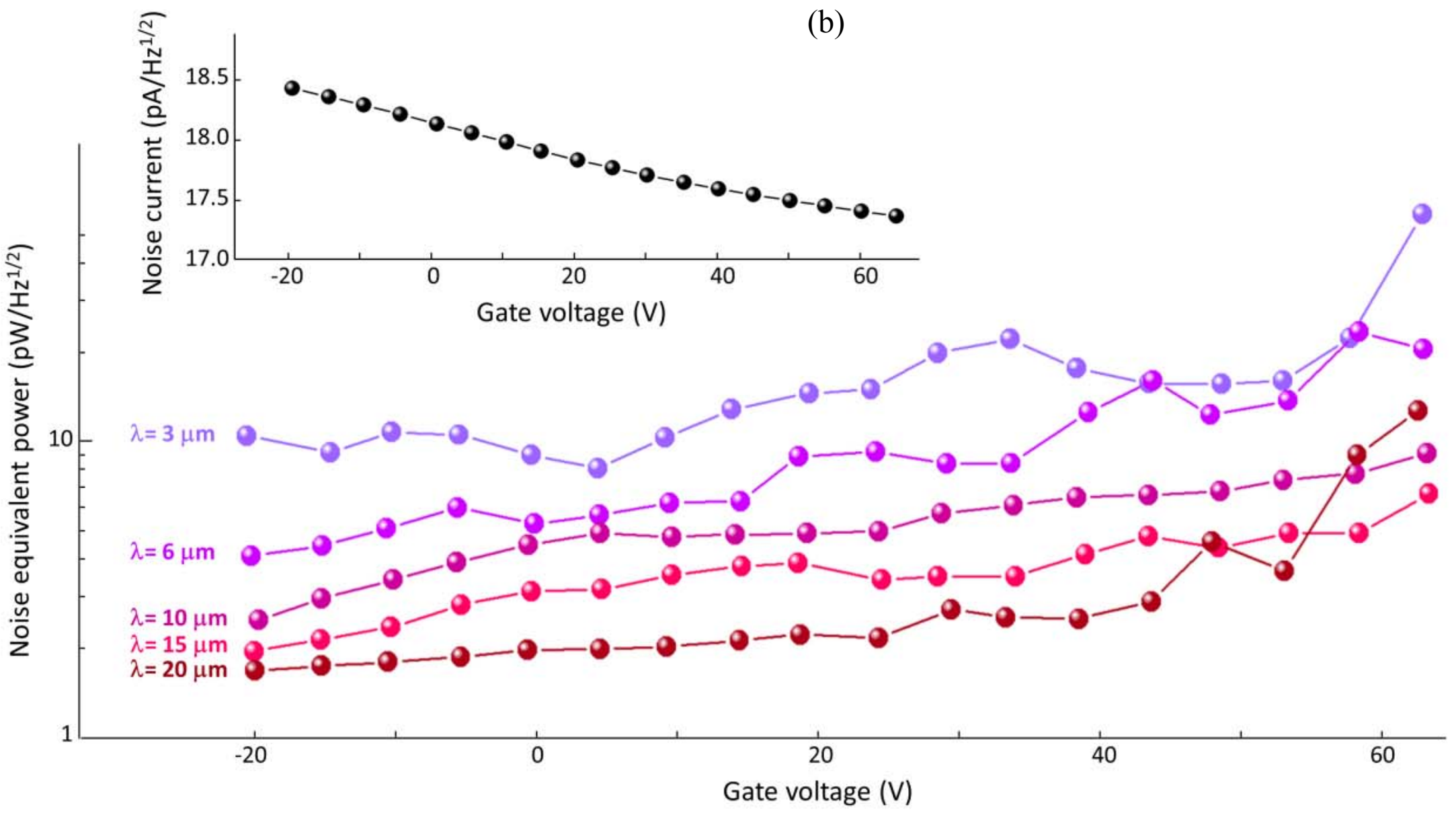

Fig. 3. Influence of the gate voltage on the photodetector performance in the infrared regime. (a) Responsivity of the fabricated photodetector at wavelengths ranging from 3 μm to 20 μm. Inset figures illustrate the intraband transitions at 3 μm and 20 μm wavelengths under gate voltages of -20 V and 65 V. The measured responsivity values have errors as large as ±20% because of thermal fluctuations in the measurement environment that lead to a ±20% variation in the output power of the Globar infrared source. (b) Noise equivalent power (NEP) of the fabricated photodetector at wavelengths ranging from 3 μm to 20 μm for an optical chopping frequency above 1 kHz. The inset shows the estimated noise current as a function of the gate voltage. All the measurements are performed at a bias voltage of $V_{bias}$ = 20 mV.

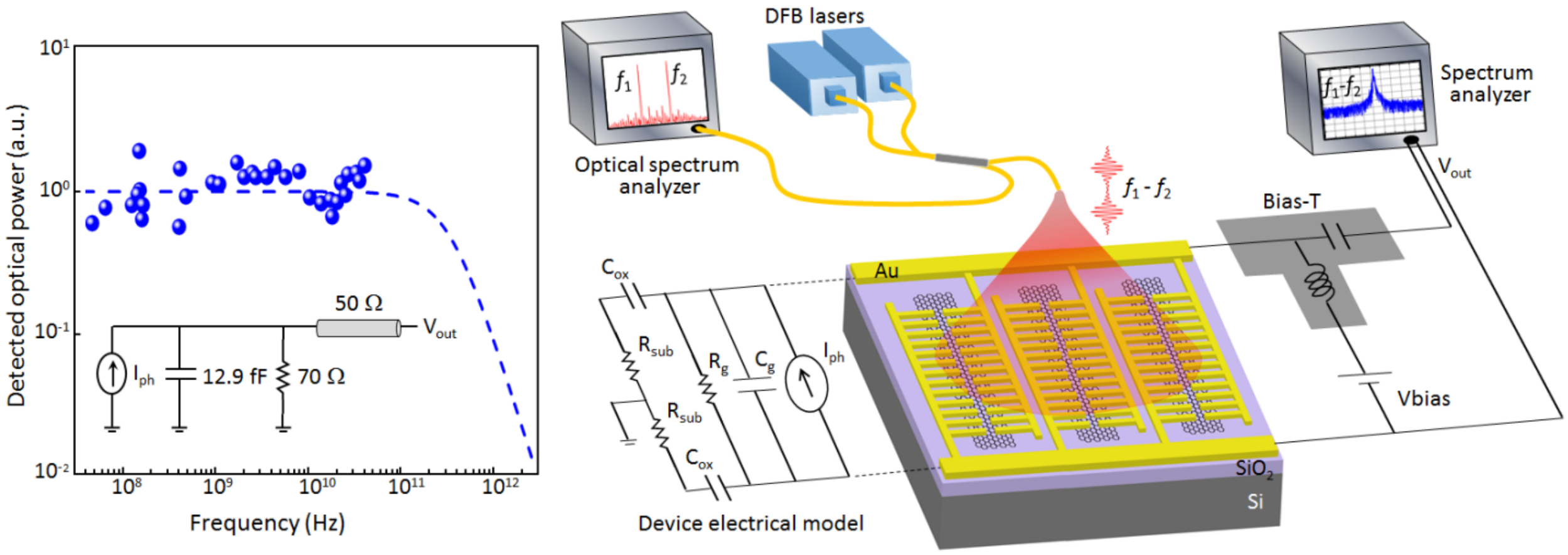

Fig. 4. Characterization of the operation speed of the fabricated photodetector. The beams from two wavelength-tunable DFB lasers at frequencies of $f_1$ and $f_2$ are focused onto the gold-patched graphene nanoribbons to generate a photocurrent, $I_{ph}$, at the optical beating frequency, $f_1 - f_2$. The measured photoresponse values exhibit no roll-off up to 50 GHz, which is the frequency limitation of the utilized experimental setup. The dashed lines show the estimated frequency response of the graphene photodetector from the device electrical model.

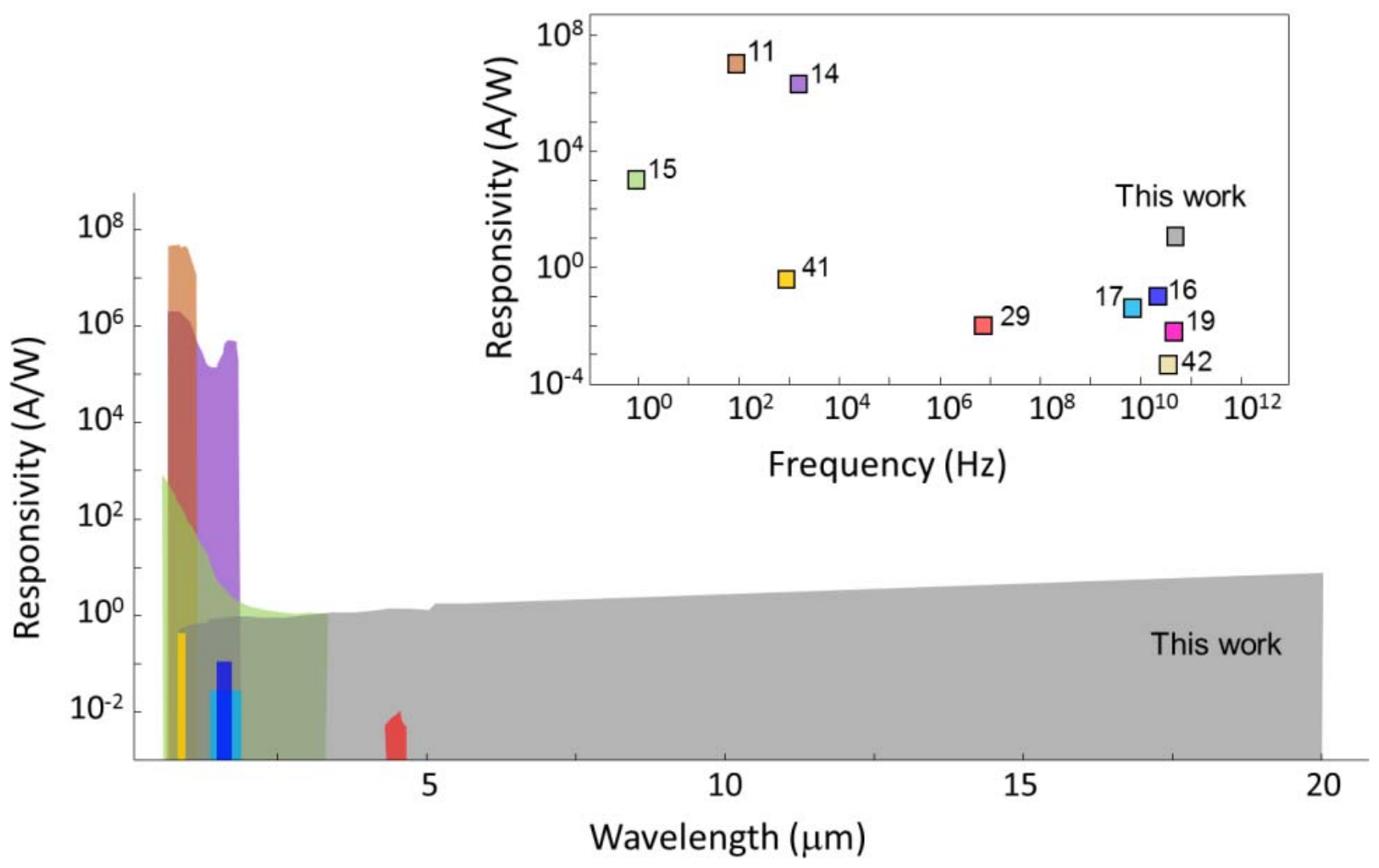

Fig. 5. Comparison of the responsivity, detection bandwidth, and operation speed of the presented room-temperature graphene photodetector with some of the highest-performance graphene photodetectors reported in the literature [11, 14, 15, 16, 17, 19, 29, 41, 42].

# Supplementary Information

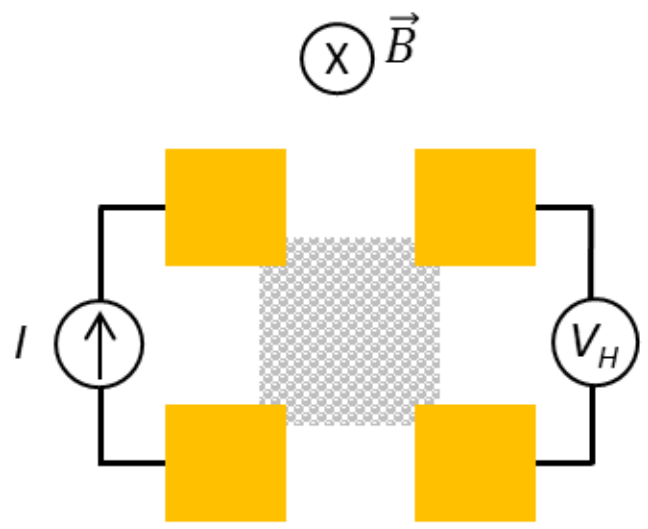

$$R_{xx} = \frac{\partial V_H}{\partial I}$$

$$R_{sheet} = \frac{\pi}{ln2} R_{xx} = 804\ \Omega/\square$$

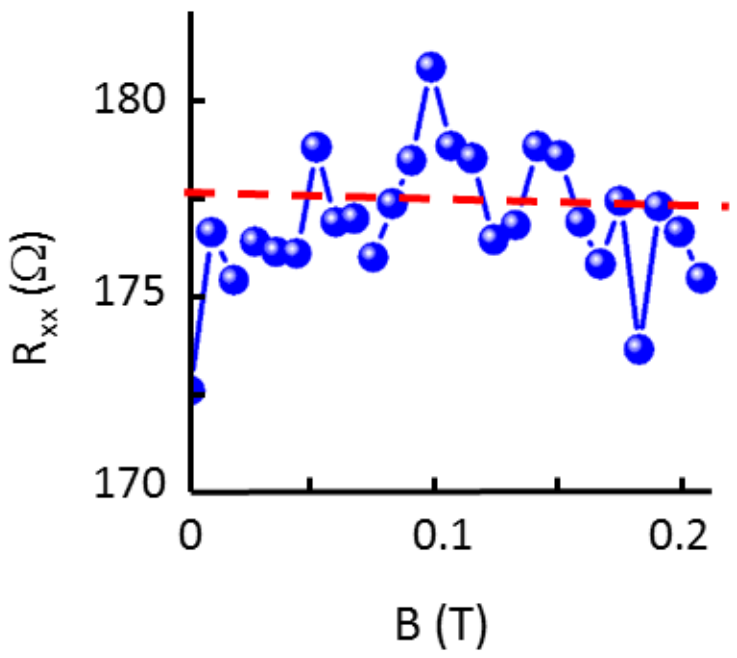

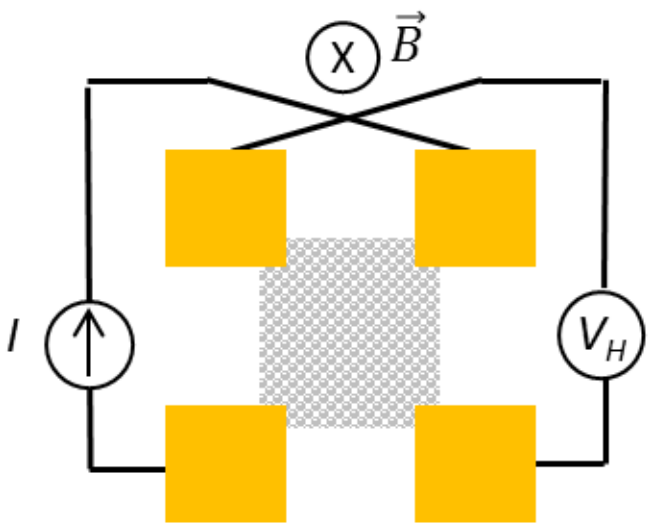

$$R_{xy} = \frac{\partial V_H}{\partial I}$$

$$n_s = \frac{1}{q} \frac{\partial B}{\partial R_{xy}} = 1.8 \times 10^{13} cm^{-2}$$

$$\mu = \frac{1}{q n_s R_s} = 432\ cm^2/Vs$$

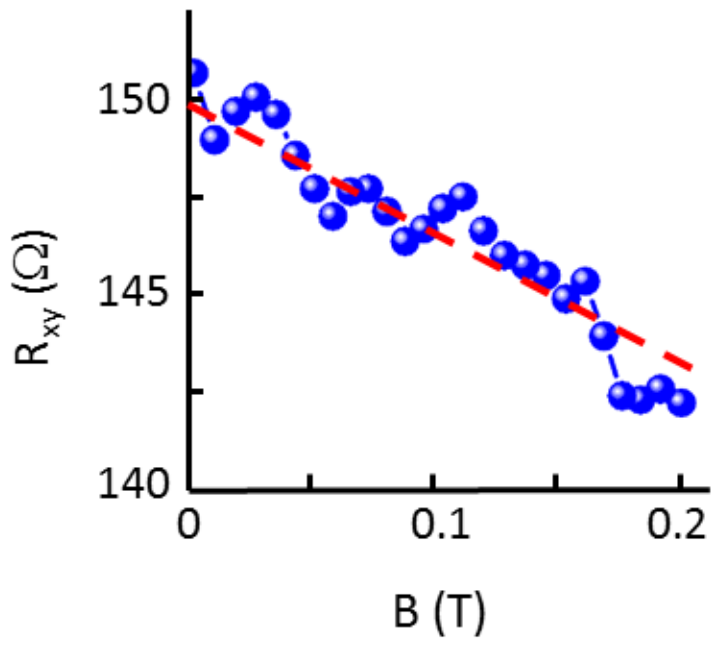

Fig. S1. Sheet resistance ($R_{sheet}$), carrier concentration ($n_s$), and mobility ($\mu$) of the utilized monolayer graphene transferred to the thermal oxide layer, measured using the van der Pauw method. Four Ti/Au contacts are used to inject current, $I$, to the graphene sheet and measure the induced Hall voltage, $V_H$, under an applied magnetic field, $B$.

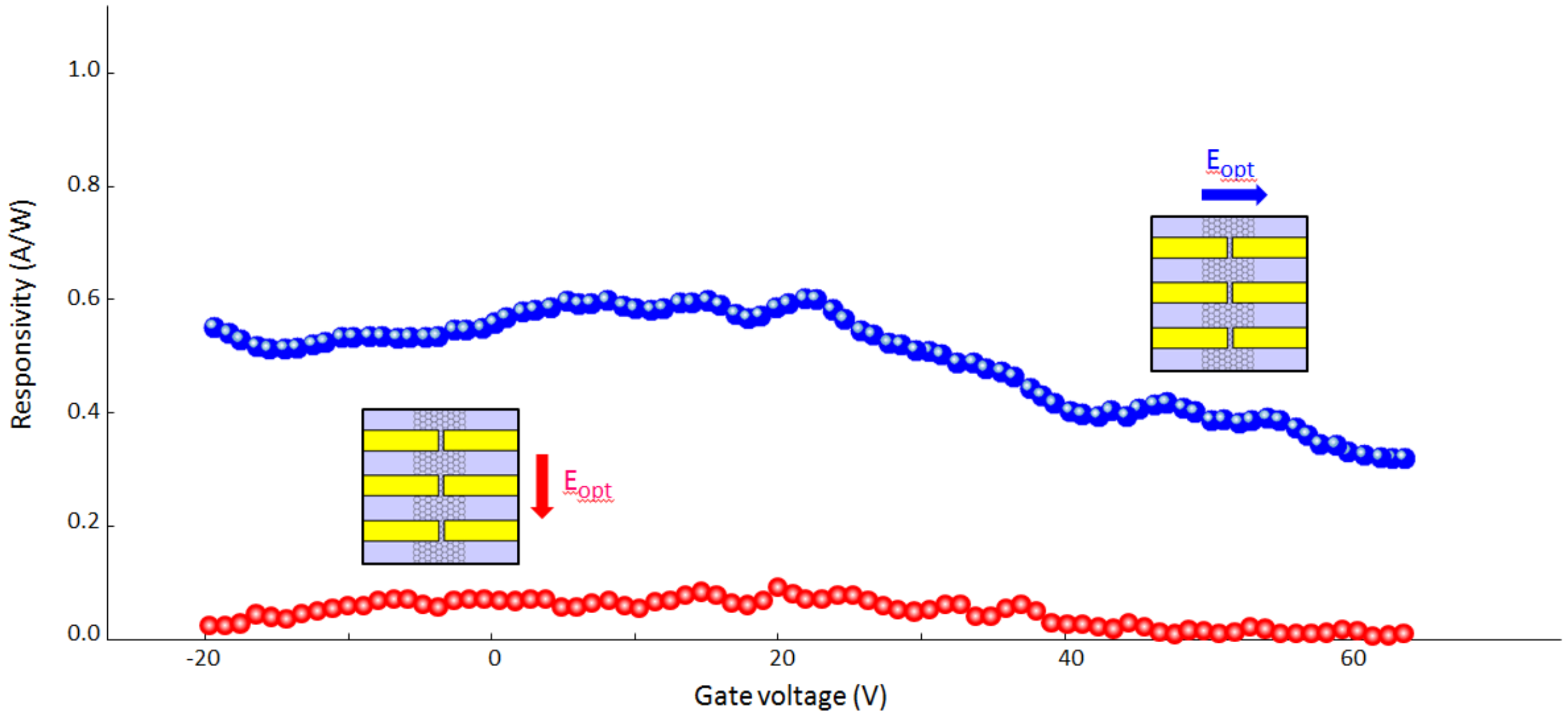

Fig. S2. Responsivity of the fabricated photodetector measured at an 800 nm wavelength under a bias voltage of 20 mV for an incident optical beam polarized normal to the graphene nanoribbons (blue data) and an incident optical beam polarized parallel with the graphene nanoribbons (red data). The responsivity data show strong polarization sensitivity due to the asymmetric geometry of the device. The strong polarization sensitivity of the presented photodetector could find many applications in polarimetric imaging and sensing systems.

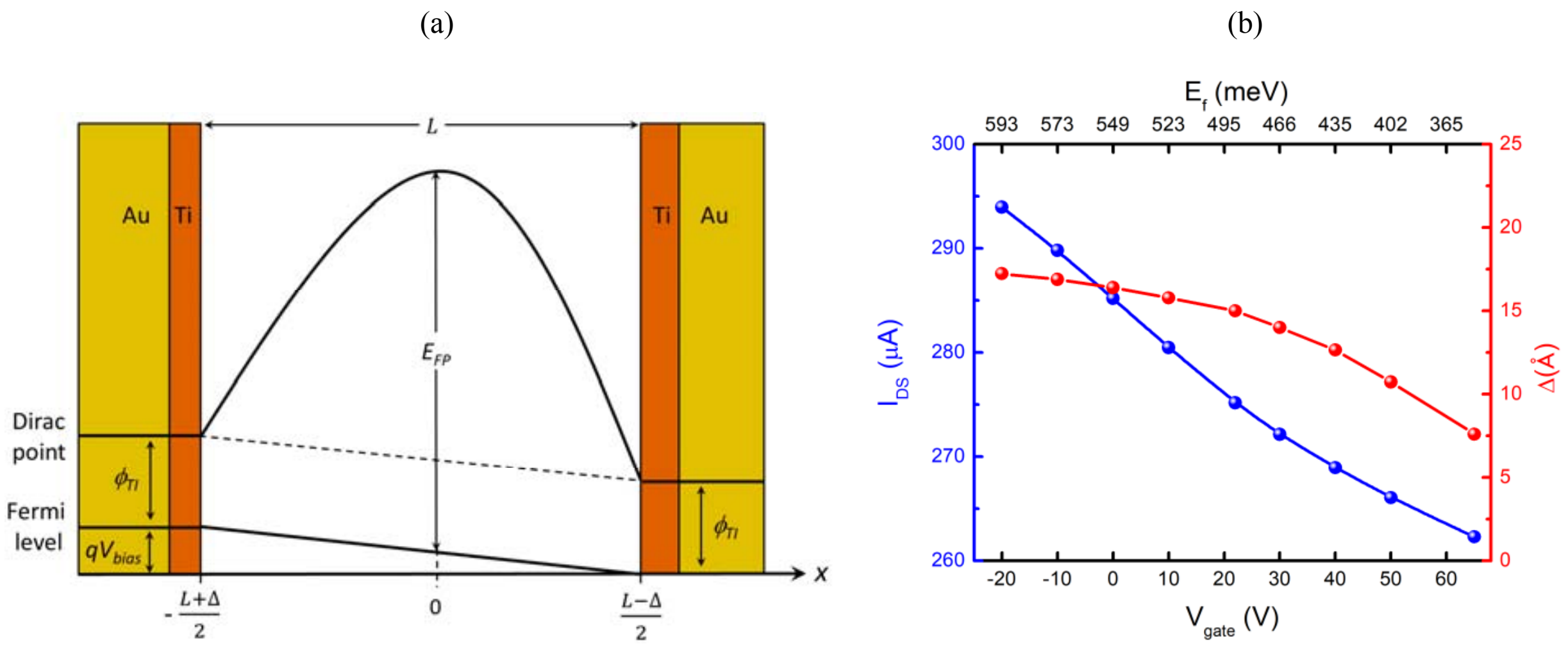

Fig. S3. (a) Band diagram of the gold-patched graphene nanoribbons: The gap between the two metal contacts to graphene ($L$ = 50 nm), which is much shorter than the transition region at the metal-graphene interface, results in a significant band-bending that dominates the device performance. The Fermi energy level of graphene at the metal contacts ($\phi_{Ti}$ = 100 meV) is determined by the graphene doping induced by adsorption on the contact surface [1]. In order to draw the device band diagram, we assume that the peak Fermi energy level of graphene ($E_{FP}$) between the metal contacts is determined by the graphene doping induced by the applied gate voltage. The validity of this assumption can be verified by the measured responsivity spectra shown in Fig. 2b. The band diagram is symmetric under a zero bias voltage with the peak Fermi energy level positioned in the middle of the gap. The potential gradient becomes steeper on the anode side under a non-zero bias voltage ($V_{bias}$) and the peak Fermi energy point is slightly shifted to the anode side by Δ. Using a second-order (quadratic) approximation, the graphene Fermi energy level between the two metal contacts can be determined as:

$$E_F(x) = \begin{cases} E_{FP} - \left(\frac{2x}{L+\Delta}\right)^2 [E_{FP} - qV_{bias} - \varphi_{Ti}] & -\frac{L+\Delta}{2} < x < 0 \\ E_{FP} - \left(\frac{2x}{L-\Delta}\right)^2 [E_{FP} - \varphi_{Ti}] & 0 < x < \frac{L-\Delta}{2} \end{cases}$$

The device dark current, which is the difference between the current at the anode contact and cathode contact, for the 8 graphene nanoribbons used in the presented photodetector is given by:

$$I_{DS} = 8I\left(\frac{L-\Delta}{2}\right) + 8I\left(-\frac{L+\Delta}{2}\right) = 8q\mu Wn(x) \times \frac{1}{q}\frac{\partial E_f(x)}{\partial x}\Big|_{x=\frac{L-\Delta}{2}} + 8q\mu Wn(x) \times \frac{1}{q}\frac{\partial E_f(x)}{\partial x}\Big|_{x=-\frac{L+\Delta}{2}}$$

where $q$ is the electron charge, $W$ is the width of the graphene nanoribbon, $\mu$ is carrier mobility, and $n(x)$ is carrier density in graphene, $n(x) = \frac{1}{\pi\hbar^2 v^2{}_F} E^2{}_F(x)$, where $v_F$ is the Fermi velocity and $\hbar$ is the Planck constant divided by 2π. (b) The measured dark current of the fabricated graphene photodetector as a function of the gate voltage at an applied bias voltage of 20 mV (blue curve) and the band diagram parameters, $E_{FP}$ and Δ, calculated from the measured dark current data at a bias voltage of 20 mV (red curve). For these calculations, we use the carrier mobility measured using the van der Pauw method (Fig. S1). The calculated parameters show a steeper potential gradient on the anode side at higher peak Fermi energy levels.

(a) (b)

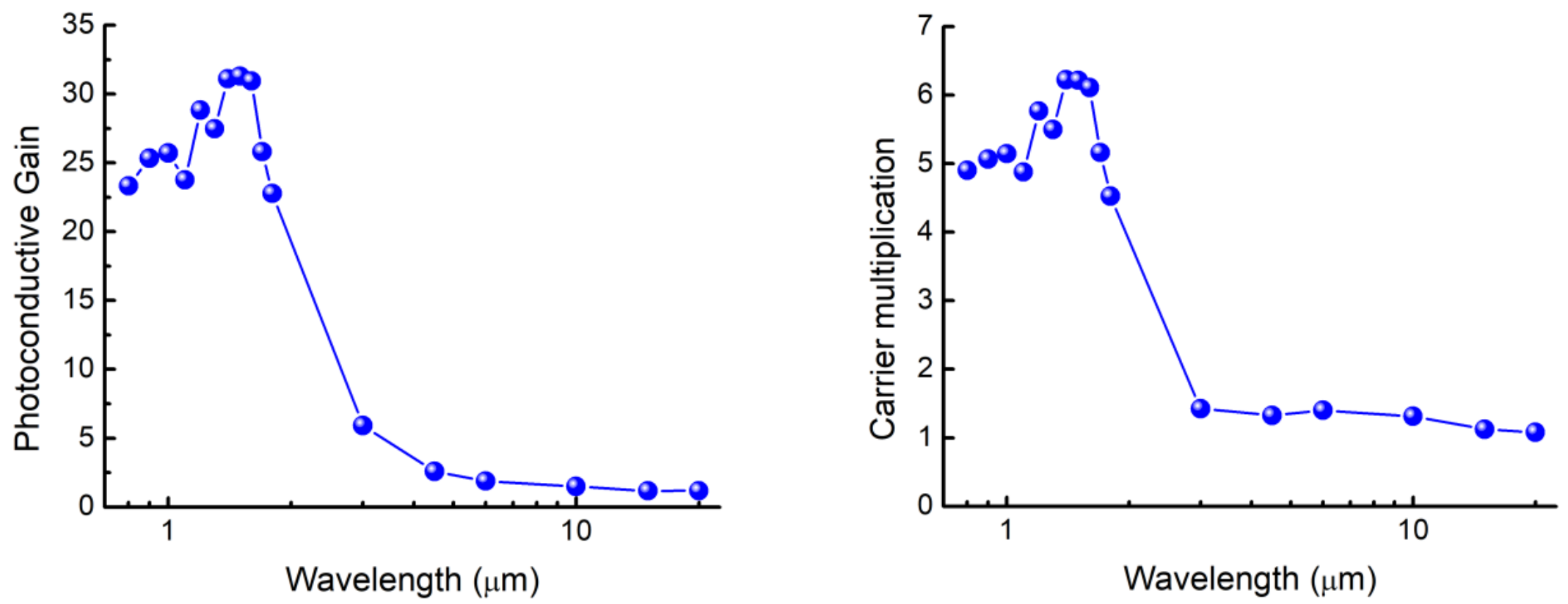

Fig. S4. (a) The measured photoconductive gain of the fabricated photodetector at an optical power of 2.5 μW, gate voltage of 22 V, and bias voltage of 20 mV. When an optical beam is incident on the device, the photogenerated electrons and holes move according to the induced electric field given by:

$$E(x) = \frac{1}{q}\frac{\partial E_F}{\partial x}$$

Therefore, the photogenerated holes move to the center of the graphene nanoribbons where they eventually recombine, the photogenerated electrons in the $\frac{L-\Delta}{2} > x > 0$ range move to the anode contact, and the photogenerated electrons in the $-\frac{L+\Delta}{2} < x < 0$ range move to the cathode contact with a carrier drift velocity given by:

$$v(x) = \mu|E(x)|/\left[1 + \frac{\mu|E(x)|}{v_{sat}}\right]$$

where $v_{sat}$ is the carrier saturation velocity. The probability for an electron in the $\frac{L-\Delta}{2} > x > 0$ range to reach the anode contact is given by $\exp(-t_a(x)/\tau_R)$, where $\tau_R$ is the photocarrier recombination lifetime and $t_a(x)$ is the carrier transit time to the anode contact:

$$t_a(x) = \int_x^{\frac{L-\Delta}{2}} \frac{dx}{v(x)}$$

Therefore, the photocurrent flowing to the anode contact is calculated as:

$$I_{ph-anode} = \left| \int_0^{\frac{L-\Delta}{2}} q \frac{\alpha P_{inc} D_p(x)}{h\nu} \frac{M}{L} e^{-\frac{t_a(x)}{\tau_R}} \, dx \right|$$

where $P_{inc}$ is the incident optical power, $D_P(x)$ is the density of the absorbed optical power inside the graphene nanoribbons calculated from the Lumerical simulations shown in Fig. 1b, $\alpha$ is the optical absorption coefficient inside the gold-patched graphene nanoribbons, $h\nu$ is photon energy, and $M$ is the carrier multiplication factor. Similarly, the probability for an electron in the $-\frac{L+\Delta}{2} < x < 0$ range to reach the cathode contact is given by $\exp(-t_c(x)/\tau_R)$, where $t_c(x)$ is the carrier transit time to the cathode contact:

$$t_c(x) = \int_{-\frac{L+\Delta}{2}}^{x} \frac{dx}{v(x)}$$

Therefore, the photocurrent flowing to the cathode contact is calculated as:

$$I_{ph-cathode} = \left| \int_0^{-\frac{L+\Delta}{2}} q \frac{\alpha P_{inc} D_P(x)}{h\nu} \frac{M}{L} e^{-\frac{t_c(x)}{\tau_R}} \, dx \right|$$

The induced photocurrent, which is the difference between the photocurrent flowing to the anode contact and cathode contact, is given by:

$$I_{ph} = I_{ph-anode} - I_{ph-cathode} = \left| \int_0^{\frac{L-\Delta}{2}} q \frac{\alpha P_{inc} D_P(x)}{h\nu} \frac{M}{L} e^{-\frac{t_a(x)}{\tau_R}} \, dx \right| - \left| \int_0^{-\frac{L+\Delta}{2}} q \frac{\alpha P_{inc} D_P(x)}{h\nu} \frac{M}{L} e^{-\frac{t_c(x)}{\tau_R}} \, dx \right|$$

By using the calculated photoconductive gain, $G$, from the measured photocurrent, $I_{ph} = qG \frac{\alpha P_{inc}}{h\nu}$, and assuming a uniform carrier multiplication factor inside the graphene nanoribbons, the carrier multiplication factor is calculated as a function of the optical wavelength.

$$M = \frac{G}{\frac{1}{L}\left| \int_0^{\frac{L-\Delta}{2}} D_P(x) e^{-\frac{t_a(x)}{\tau_R}} \, dx \right| - \frac{1}{L}\left| \int_0^{-\frac{L+\Delta}{2}} D_P(x) e^{-\frac{t_c(x)}{\tau_R}} \, dx \right|}$$

(b) The estimated carrier multiplication factor at an optical power of 2.5 µW, gate voltage of 22 V, and bias voltage of 20 mV for a carrier saturation velocity of $2\times10^7$ cm/s [2] and a photocarrier recombination lifetime of 1 ps [3-5] as a function of the optical wavelength. The results indicate carrier multiplication factors larger than 1 in the visible and near-infrared frequencies (smaller carrier saturation velocity values down to $0.1\times10^7$ cm/s would only reduce the estimated carrier multiplication factors by 10%). There are two possible mechanisms for carrier multiplication. The first mechanism involves an interband scattering process (also called impact ionization) in which an excited electron in the conduction band relaxes to a lower energy state, and the released energy is transferred for the excitation of another electron from the valance band to the conduction band. This mechanism increases the carrier density within the conduction band and is suppressed by the Pauli blocking. Therefore, increasing graphene doping reduces its efficiency [6-8]. The second mechanism involves a Coulomb-induced intraband scattering process (also called hot carrier multiplication) in which an excited electron relaxes to a lower energy state, and the released energy is transferred for the excitation of another electron below the Fermi energy level to a state above the Fermi energy level. This mechanism becomes more efficient at higher doping levels due to the availability of a larger number/density of states that electrons can be excited to [6, 7, 9]. Considering the relatively high carrier concentration levels in the graphene nanoribbons, the observed carrier multiplication in the presented graphene photodetector is anticipated to be governed by the second mechanism, the Coulomb-induced intraband scattering process. As expected, higher carrier multiplication factors are achieved in the visible and near-infrared wavelengths compared to the infrared wavelengths. The reason is that the photogenerated electrons in response to the visible and near-infrared beams are excited to higher energy levels in the conduction band. As a result, they give rise to the excitation of a larger number of secondary electrons by transferring more energy during relaxation. Also, lower carrier multiplication factors are achieved at higher optical power levels. This decrease is due to the increase in the carrier recombination rate at high photogenerated carrier densities, which reduces the carrier scattering time and the carrier multiplication efficiency [8-10].

As it can be observed from the above discussions, only the photovoltaic (PV) response of the gold-patched graphene nanoribbons has been considered to estimate the carrier multiplication factor in graphene. Our calculations show that the photothermoelectric (PTE) effect does not play a significant role in the photo-response of the presented photodetector due to the geometric symmetry of the gold-patched graphene nanoribbons, which offers a relatively symmetric profile of the density of states and temperature for the PTE-induced photocurrents flowing from graphene to the metal contacts. Including the PTE-induced photocurrent in our calculations would only slightly increase the estimated carrier multiplication factor. This is because the direction of the PTE-induced photocurrent in the presented photodetector is in the opposite direction to the PV-induced photocurrent. The reason is that hot carriers diffuse from regions with lower density of states to regions with higher density of states to maximize the entropy. Since the Fermi energy level has a slightly steeper slope on the anode junction than the cathode junction, the PTE-induced electron current flowing from the anode junction (lower density of states) to the graphene nanoribbons (higher density of states) to maximize the entropy would be larger than the PTE-induced electron current flowing from the cathode junction (lower density of states) to the graphene nanoribbons (higher density of states) to maximize the entropy.

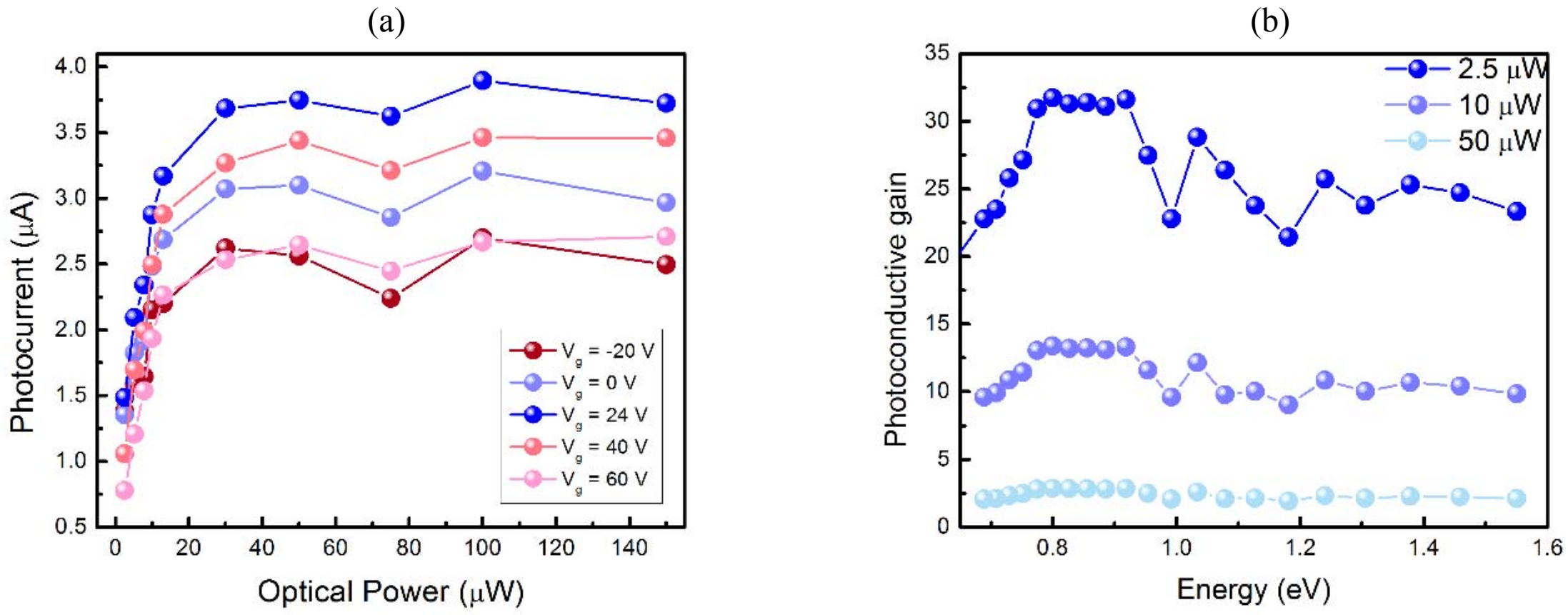

Fig. S5. (a) Output photocurrent of the fabricated graphene photodetector at an 800 nm wavelength as a function of the incident optical power and gate voltage. (b) Photoconductive gain of the fabricated graphene photodetector in the visible and near-infrared wavelength ranges as a function of the incident optical power. All of the measurements are performed at an applied bias voltage of 20 mV.

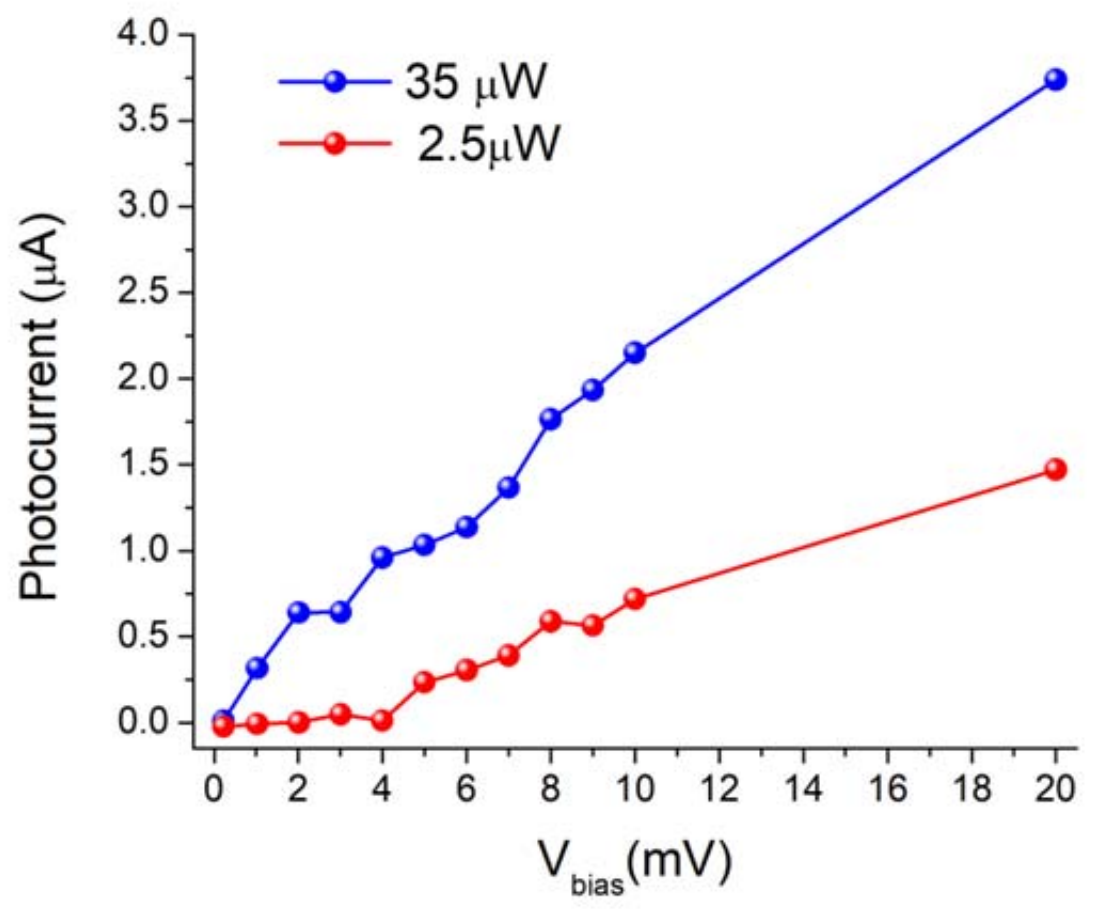

Fig. S6. Output photocurrent of the fabricated graphene photodetector at an 800 nm wavelength as a function of the applied bias voltage for an optical power of 2.5 μW (red data) and 35 μW (blue data).

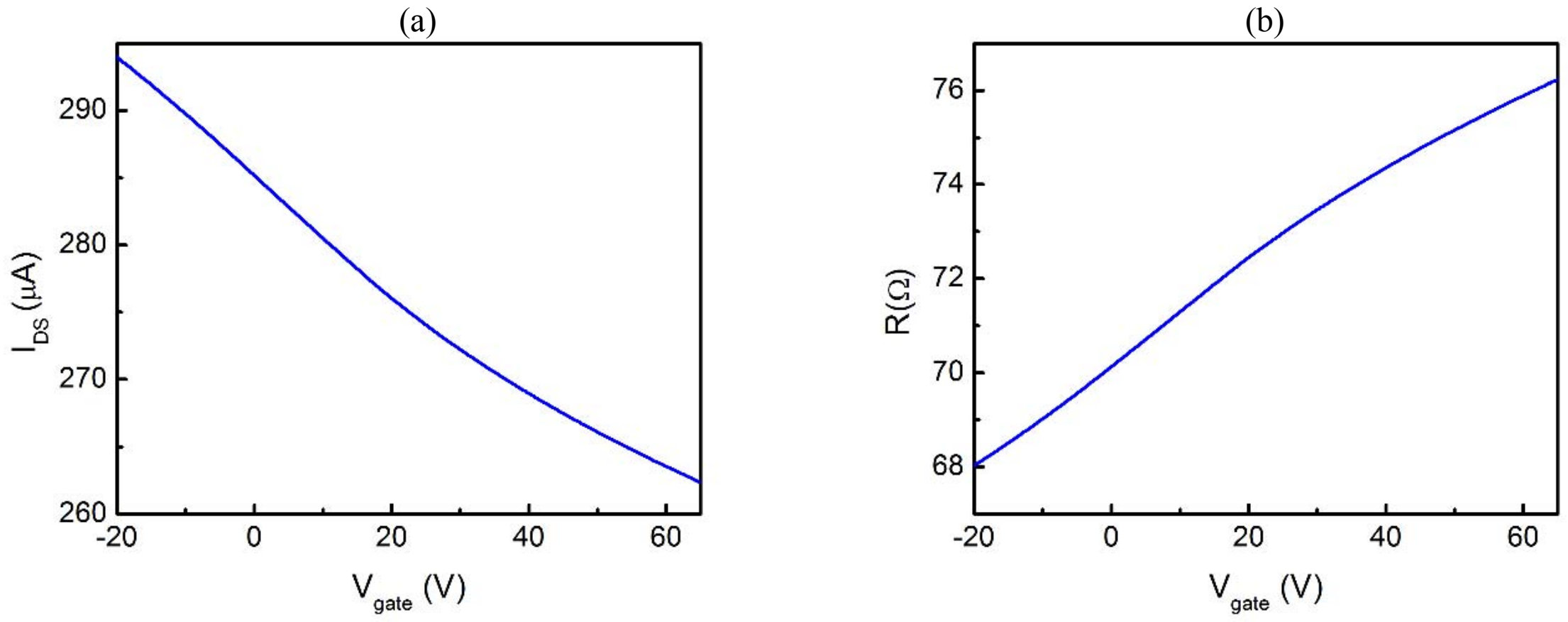

Fig. S7. (a) Dark current of the fabricated graphene photodetector as a function of the gate voltage at an applied bias voltage of 20 mV. (b) Resistance of the fabricated graphene photodetector as a function of the gate voltage.

| Model number | Center wavelength | Power transmission | Filter bandwidth |
|---|---|---|---|
| Thorlabs FB3000-500 | 3 μm | 1.08 % | 0.48 μm |
| Thorlabs FB4500-500 | 4.5 μm | 0.98 % | 0.51 μm |
| Thorlabs FB6000-500 | 6 μm | 0.79 % | 0.38 μm |
| Spectrogon BBP9900-11000 | 10 μm | 0.56 % | 1.7 μm |
| Thorlabs FB19M15 | 15 μm | 0.94 % | 3.1 μm |
| Thorlabs FB19M20 | 20 μm | 1.06 % | 2.8 μm |

Table S1. Characteristics of the utilized infrared filters for the infrared responsivity measurements.